\documentclass[apl,graphicx, reprint, twocolumn,superscriptaddress]{revtex4-1}

\usepackage{graphicx}
\usepackage{natbib}
\usepackage{amsmath}
\usepackage{color}
\draft % marks overfull lines with a black rule on the right

\begin{document}

%\preprint{}

%\title{Magnetism in a single CrO$_2$ micro-grain studied by Hall magnetometry technique} %Title of paper
\title{On the influence of nanometer-thin antiferromagnetic surface layer on ferromagnetic CrO$_2$ %/ Unusual low-temperature magnetic anomaly in a single micro-crystal of CrO$_2$ studied by micro-Hall magnetometry
 }



\author{P. Das}
%\email[]{das@physik.uni-frankfurt.de}
%\thanks{}
%\altaffiliation{Max Planck Institute for Chemical Physics of Solids, Noethnitzer Str. 40, 01187 Dresden, Germany}
\affiliation{Institute of Physics, Goethe University, Max von Laue Str. 1, 60438 Frankfurt (M), Germany.}
\affiliation{Max Planck Institute for Chemical Physics of Solids, Noethnitzer Str. 40, 01187 Dresden, Germany}

\author{A. Bajpai}
%\email[]{das@physik.uni-frankfurt.de}
%\thanks{}
\affiliation{Indian Institute of Science Education and Research, Pashan Road, 411021 Pune, India}
%\affiliation{Institute of Solid State Research, IFW-Dresden, 01171 Dresden, Germany.}

%\author{F. Porrati}
%\email[]{das@physik.uni-frankfurt.de}
%\thanks{}
%\altaffiliation{Max Planck Institute for Chemical Physics of Solids, Noethnitzer Str. 40, 01187 Dresden, Germany}
%\affiliation{Institute of Physics, Goethe University, Max von Laue Str. 1, 60438 Frankfurt (M), Germany.}

%\author{S. Wirth}
%\email[]{das@physik.uni-frankfurt.de}
%\thanks{}
%\altaffiliation{Max Planck Institute for Chemical Physics of Solids, Noethnitzer Str. 40, 01187 Dresden, Germany}
%\affiliation{Max Planck Institute for Chemical Physics of Solids, Noethnitzer Str. 40, 01187 Dresden, Germany.}

%\author{M. Huth}
%\email[]{das@physik.uni-frankfurt.de}
%\thanks{}
%\altaffiliation{Max Planck Institute for Chemical Physics of Solids, Noethnitzer Str. 40, 01187 Dresden, Germany}
%\affiliation{Institute of Physics, Goethe University, Max von Laue Str. 1, 60438 Frankfurt (M), Germany.}

\author{Y. Ohno}
%\email[]{das@physik.uni-frankfurt.de}
%\thanks{}
%\altaffiliation{Max Planck Institute for Chemical Physics of Solids, Noethnitzer Str. 40, 01187 Dresden, Germany}
\affiliation{Laboratory for Nanoelectronics and Spintronics, Research Institute of Electrical Communication, Tohoku University, Sendai 980-8577, Japan.}

\author{H. Ohno}
%\email[]{das@physik.uni-frankfurt.de}
%\thanks{}
%\altaffiliation{Max Planck Institute for Chemical Physics of Solids, Noethnitzer Str. 40, 01187 Dresden, Germany}
\affiliation{Laboratory for Nanoelectronics and Spintronics, Research Institute of Electrical Communication, Tohoku University, Sendai 980-8577, Japan.}

\author{J. M\"uller}
%\email[]{das@physik.uni-frankfurt.de}
%\thanks{}
%\altaffiliation{Max Planck Institute for Chemical Physics of Solids, Noethnitzer Str. 40, 01187 Dresden, Germany}
\affiliation{Institute of Physics, Goethe University, Max von Laue Str. 1, 60438 Frankfurt (M), Germany.}

%\collaboration{F. Porrati}

 \date{\today}

\begin{abstract}

We present magnetic stray field measurements performed on a single micro-crystal of the half metallic ferromagnet CrO$_2$, covered by a naturally grown 2\,-\,5\,nm surface layer of antiferromagnetic (AFM) Cr$_2$O$_3$. The temperature variation of the stray field of the micro-crystal measured by micro-Hall magnetometry shows an anomalous increase below $\sim$\,60\,K. We find clear evidence that this behavior is due to the influence of the  AFM surface layer, which could not be isolated in the corresponding bulk magnetization data measured using SQUID magnetometry. The distribution of pinning potentials, analyzed from Barkhausen jumps, exhibits a similar temperature dependence. Overall, the results indicate that the surface layer plays a role in defining the potential landscape seen by the domain configuration in the ferromagnetic grain.
\end{abstract}

\pacs{}% insert suggested PACS numbers in braces on next line

\maketitle %\maketitle must follow title, authors, abstract and \pacs


%\section{
 In the modern age of fast and miniaturized electronics, as Moore's law is set to be rattled \cite{LothScience2012}, the use of the electron's spin degree of freedom in multifunctional devices has been the focus of current research as it is expected to lead to a dramatic improvement of device capability and performance. Both the control and manipulation of electron spins and search for new materials with high spin polarization, preferably at room temperature (RT), have attracted considerable interest of researchers in recent years. Apart from dilute ferromagnetic (FM) semiconductors where the Curie temperature is still below RT \cite{OhnoNatMat2010}, the search is on to discover other new materials with high spin polarization, improved spin life times, etc.~\cite{OpelJPD2012}. The binary oxide material CrO$_2$ is well known in data storage technology where it has been used successfully in magnetic recording tapes. The material, which is a half-metallic ferromagnet with $T_{\rm{C}}$\,$\approx$\,395\,K, has attracted renewed interest as an important material for spintronics applications as it exhibits spin polarization of $\sim\,100$\% at low temperatures~\cite{SoulenScience1998, WolfScience2001}. In the recent years, several attempts have been made to understand the electronic transport and magnetic behavior of this material, see, e.g., Refs.~\cite{HwangScience1997, CoeyPRL1998, LiAPL1999, WattsPRB2000, GoeringPRL2002, MiaoJAP2005, BajpaiPRB2007, DasAPL2010}. It is well known that the metastable character of CrO$_2$ leads to surface decomposition forming Cr$_2$O$_3$ or Cr$_2$O$_5$, which are electrically insulating and AFM with bulk $T_{\rm{N}}$\,$\sim$\,307\,K and $\,\sim$\,125\,K, respectively~\cite{McGuirePR1956}. It has been observed that polycrystalline
 CrO$_2$ material with such insulating AFM grain boundaries exhibit enhanced low-field magnetoresistance (MR) properties compared to single crystals of pure CrO$_2$~\cite{HwangScience1997, BajpaiPRB2007}. Recently, by using a different route for synthesis, Bajpai \textit{et al.} have synthesized CrO$_2$ grains of larger size as compared to the commercially available ones exhibiting much improved MR close to RT~\cite{BajpaiAPL2005, BajpaiPRB2007, BajpaiPatent}. It was observed that the MR of such polycrystalline samples is sensitive to the nature of the insulating surface magnetic layer~\cite{BajpaiPRB2007}. Thus, a detailed understanding of the magnetotransport behavior of polycrystalline CrO$_2$ requires a thorough investigation of its magnetic properties which may be influenced by the AFM surface layer.

\begin{figure}[b]
\begin{center}
\includegraphics[width=0.45\textwidth]{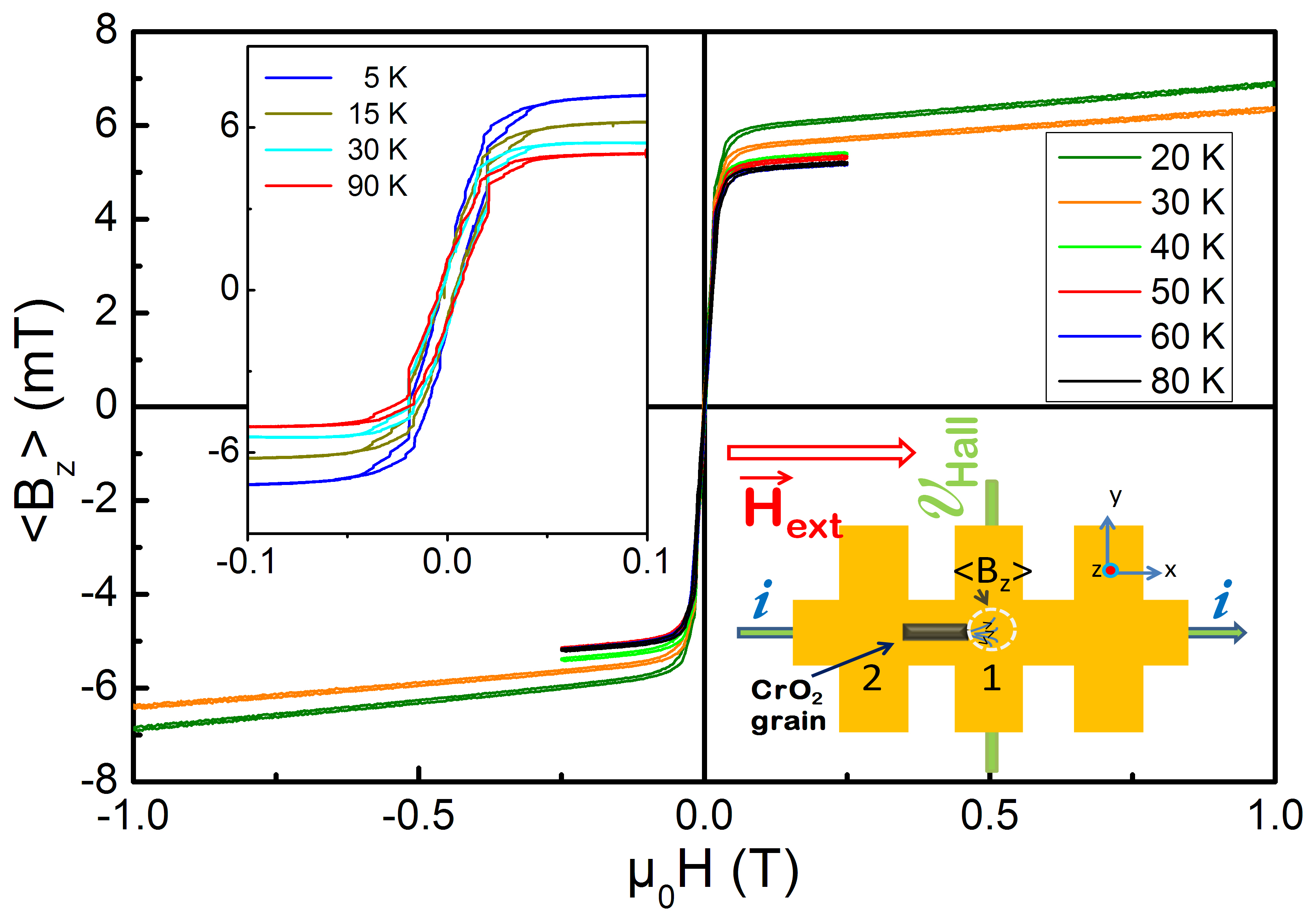}
\caption{\label{Fig_Hyst_T}(Color online) Hysteresis loops measured at end 1 (see schematic diagram of the  set-up in right inset) of the CrO$_2$ micro-crystal at different temperatures for $\mu_0H_{\rm{ext}}$ parallel to the EMD. The measurements were carried out at an external field sweep rate of 50\,mT/min. Shown in the left inset are a set of data obtained by sweeping the field at a slow rate of 1\,mT/min, where stochastic Barkhausen jumps, in addition to a non-stochastic jump at $\mu_0H_{\rm{ext}}$\,=\,20\,mT, are clearly visible. See Ref.\cite{DasAPL2010} for details. }
\end{center}
\end{figure}

 Such CrO$_2$ micro-crystals have shown many novel effects arising from the FM/AFM interface, such as enhanced magnetoresistance, a strong temperature dependent magnetoelectric effect, an unusual field dependence of remanent magnetization, etc. \cite{CoeyPRL1998, BajpaiEPL2010, BajpaiJPCM2010}. However, it should be noted that all these measurements were performed on bulk pellets consisting of an assembly of such micro-crystals (or grains) oriented in random directions. While bulk magnetization measurements on such pellets were performed using conventional SQUID magnetometers, micro-Hall magnetometry on a single micro-crystal picked from the same pellet provides a unique opportunity to study and compare the magnetization mechanism (domain wall dynamics) of a bulk sample to its constituent single entities. The results on a micro-crystal do not suffer from averaging effects. Therefore, in this work we have aimed for investigating a \emph{single} micro-crystal of CrO$_2$, with a thin polycrystalline Cr$_2$O$_3$ surface layer. We find a pronounced increase of the sample's magnetic signal at low temperatures ($T\,\ll\,T_C$), which we demonstrate being an effect of magnetic interaction with the AFM Cr$_2$O$_3$ surface layer.

The single CrO$_2$ micro-crystal used in the present study was picked from a granular sample, synthesized by the procedure described in Ref.~\cite{BajpaiAPL2005}. The micro-crystal has dimensions of approximately 7$\times$1.2$\times$0.5\,$\mu$m$^3$ and has a 2\,-\,5\,nm thin surface layer of Cr$_2$O$_3$ which is polycrystalline in nature, as seen in Transmission Electron Microscopy measurements, reported in \cite{BajpaiEPL2010}. The measurements of pure Cr$_2$O$_3$ were performed on polycrystalline powders procured from Sigma Aldrich Co.
\begin{figure}[t]
\begin{center}
\includegraphics[width=0.45\textwidth]{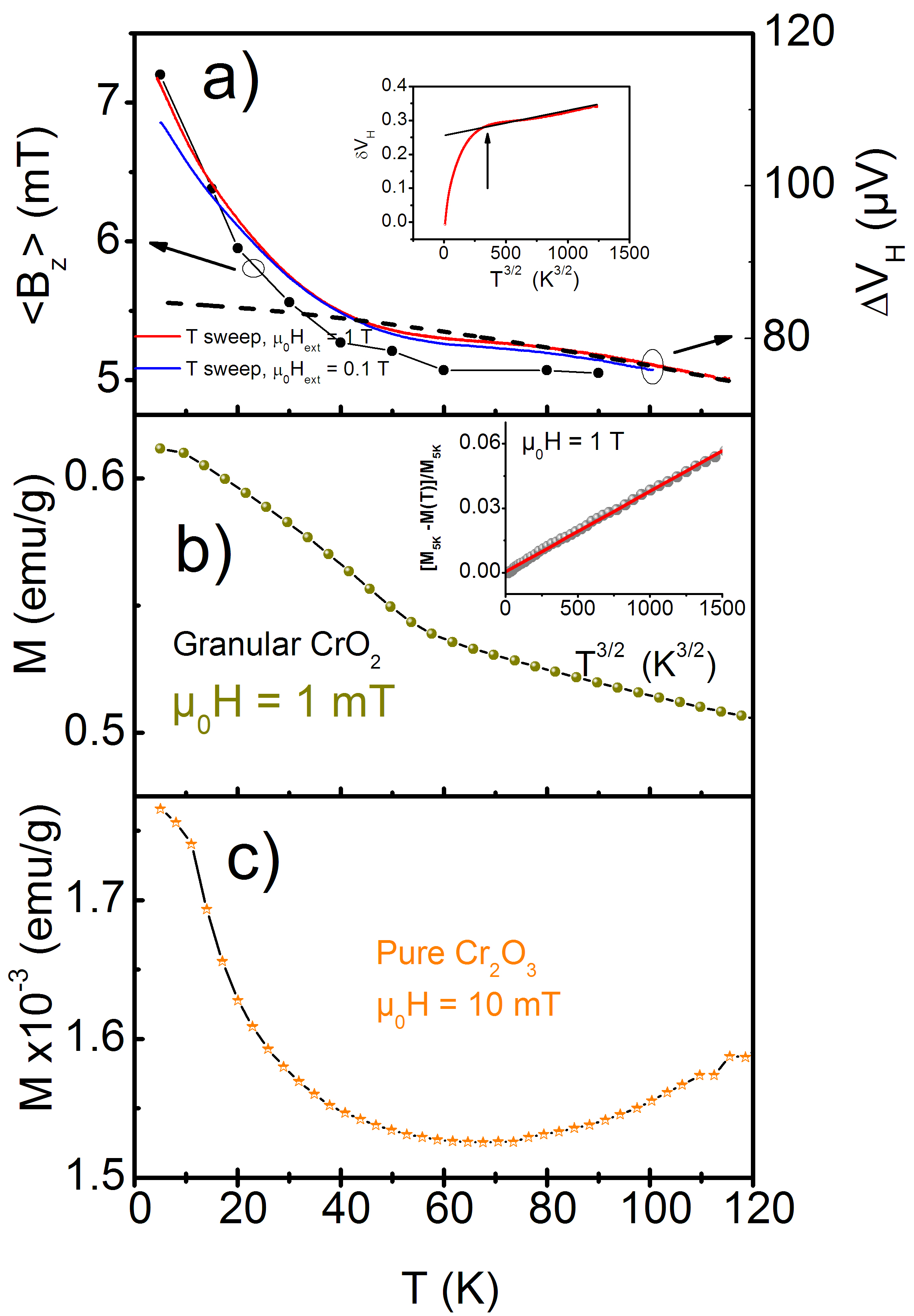} 
\caption{\label{Fig_Satfield_T} (Color online) (a) Magnetic stray field (from micro-Hall magnetometry) of the CrO$_2$ micro-crystal at different $T$ obtained from Fig.~\ref{Fig_Hyst_T} (measured at $\mu_0H_{ext}$\,=\,0.1\,T) and continuous temperature sweep measurements at $\mu_0H_{\rm{ext}}$\,=\,0.1\,T and $\mu_0H_{\rm{ext}}$\,=\,1\,T, respectively. The dashed line shows the derived Bloch's $T^{3/2}$ behavior. The deviation from Bloch's $T^{3/2}$ law is shown clearly in the inset which shows $\delta V_H$ (=$\frac{\Delta V_H(5\,K)-\Delta V_H(T)}{\Delta V_H(5\,K)}$). The straight line in the inset is a linear fit of $\delta V_H$ \emph{vs.} $T^{3/2}$ data. (b) SQUID measurements of $M$ $\mathit{vs.}$  $T$ of a bulk pellet (from which the single grain was picked) showing a feature at around 60\,K for an external field of 1\,mT. The magnetization at large field follows Bloch's $T^{3/2}$ law, as shown in the inset. (c) $M$ \textit{vs.} $T$ for pure Cr$_2$O$_3$ powder showing a minimum at $\sim$\,60\,K.}
\end{center}
\end{figure}
For the magnetic measurements of the rod-shaped CrO$_2$ micro-crystal, the stray field emanating from one end of the crystal was measured utilizing a micro-Hall device based on a two-dimensional electron gas (2DEG) formed at the interface of a GaAs/AlGaAs heterostructure. Hall crosses of dimensions 5$\,\times\,5\,\mu m^2$ were prepared using standard photolithography followed by chemical wet etching \cite{DasAPL2010}.
The measurements were performed with the external magnetic field ($\mu_0H_{\rm{ext}}$) applied parallel to the current (for details, see the schematic diagram in the inset of Fig.\,\ref{Fig_Hyst_T}) so that the measured Hall voltage is only due to the $z$-component of the average stray field $\langle B_z\rangle$ emanating from the sample. For the CrO$_2$ micro-crystal, this direction is along the long axis, which is also the easy magnetization direction (EMD), i.e., along the [001] crystallographic direction. Thus, the total anisotropy in the present case consists of both shape and magnetocrystalline anisotropy. A small contribution of the external field due to the misalignment (angle $\sim\,2^{\circ}$) of the field with respect to the plane of the Hall-device can be accounted for by measuring an empty reference cross. The measured Hall voltage $\Delta V_H$, where the background has been subtracted, can be converted to the stray field $\langle B_z\rangle $ averaged over the active area of the Hall sensor from the knowledge of the Hall coefficient at the corresponding temperatures. Our previously published analysis of the magnetic state and domain configuration of the micro-crystal showed that the magnetization reversal takes place due to the motion of a \emph{single} cross-tie domain wall~\cite{DasAPL2010, DasJPCS2011}. The measurements on bulk pellets of granular CrO$_2$ and pure Cr$_2$O$_3$, respectively, were carried out using a commercial SQUID magnetometer.

 Figure~\ref{Fig_Hyst_T} shows the $\langle B_z\rangle$ \emph{vs.} $H_{\rm{ext}}$ isotherms of the CrO$_2$ micro-crystal for temperatures 20\,K\,-\,80\,K. For two temperatures, the field is shown up to 1\,T, the ideal saturation for the pure CrO$_2$ phase~\cite{CoeyPRL1998, BajpaiPRB2007}. The data show an unusual temperature dependence of $\langle B_z\rangle $ for $T\,\ll\,T_{\rm{C}}$ with $\langle B_z\rangle$ rapidly increasing at low temperatures. In addition to this, we also observe that $\langle B_z\rangle$ at lower temperatures exhibits a finite slope (indicating non-saturation) in the otherwise saturated regime. Note that the data shown in Fig.~\ref{Fig_Hyst_T} were corrected for background Hall voltage which is linear in field. Thus, the slope in $\langle B_z\rangle$, in the saturated regime of pure CrO$_2$ appears to be a characteristic for the micro-crystal.

\begin{figure}[t]
\begin{center}
\includegraphics[width=0.5\textwidth]{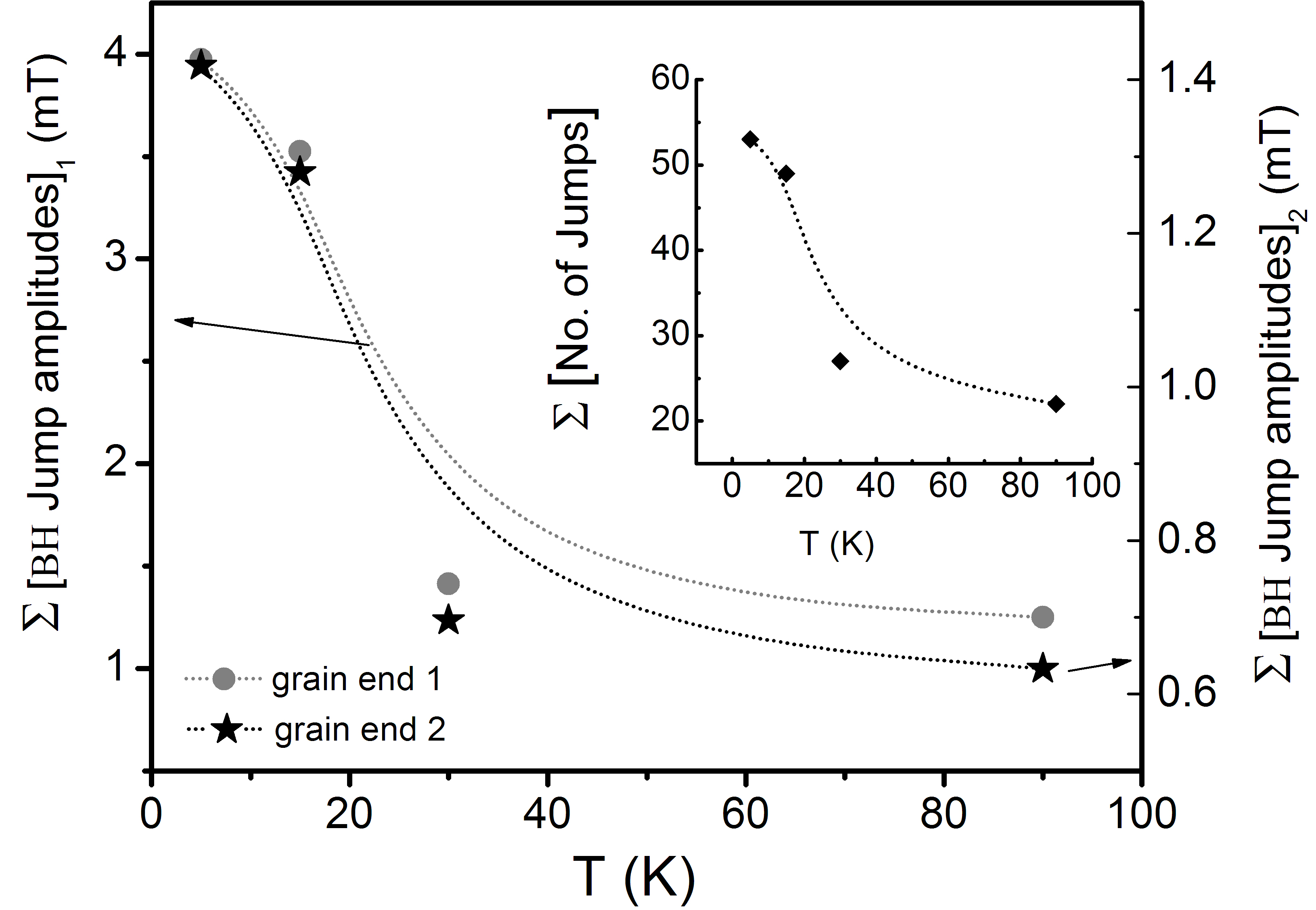}
\caption{\label{Fig_BH-jumps1} Amplitudes of Barkhausen signal (sum of the amplitudes of all individual Barkhausen jumps at the two ends of the micro-crystal) observed at different temperatures. Inset shows the dependence of total number of Barkhausen jumps observed at one grain end on temperature. The dotted lines are guides to the eyes.}
\end{center}
\end{figure}

To further understand this behavior of the micro-crystal, we have carried out continuous temperature sweeps while measuring the Hall signal for $\mu_0H_{\rm{ext}}$\,=\,0.1\,T and 1\,T. In this case also, a simultaneous measurement of the background Hall signal was carried out on an empty reference cross which was subtracted from the measured data. Figure~\ref{Fig_Satfield_T}(a) shows the temperature variation of this background corrected Hall signal. For both values of the external field, we observe a clear increase of the magnetic signal deviating from Bloch's $T^{3/2}$ behavior (shown by the dashed line) below about 60\,K. Surprisingly, this anomaly is not observed in the corresponding SQUID measurements for bulk samples (in pellet form), when measured in the field range of 1\,T (cf. inset of Fig.~\ref{Fig_Satfield_T}(b)). However, as shown in Fig.~\ref{Fig_Satfield_T}(b), measurements in a very low field of 1\,mT exhibit a similar anomaly at $\sim$\,60\,K. The effect appears to be strongly history dependent. We observe that for the bulk sample, the magnetization as a function of temperature exhibits Bloch's $T^{3/2}$ behavior (see inset of Fig.~\ref{Fig_Satfield_T}(b)) in the above-mentioned temperature range, for all fields above $\sim$\,5\,mT~\cite{BajpaiPRB2007} in contrast to the data obtained from a \emph{single} micro-crystal.

Measurements of pure Cr$_2$O$_3$ exhibit a similar increase of the magnetization $M$ at low temperatures (see Fig.~\ref{Fig_Satfield_T}(c)). Thus, we find a qualitatively similar functional form of the magnetic signal at low temperatures for both the single CrO$_2$ micro-crystal with Cr$_2$O$_3$ surface layer and pure Cr$_2$O$_3$ phase. The comparison of the corresponding experimental data indicates that the functional form of CrO$_2$ at low temperatures is influenced by the presence of the AFM Cr$_2$O$_3$ surface layer. While this effect is subtle and is seen only in the very low field regime for randomly-oriented grains in bulk pellets, the measurements on \emph{single} grain reveal the influence of the AFM layer which is robust in nature up in fields of 1\,T. As the coercivity of CrO$_2$ is $\sim$\,1.5\,-\,5\,mT~\cite{BajpaiAPL2005, DasAPL2010, LiAPL1999}, $T$-dependent bulk measurements of pellets using SQUID in this small field range may show the effective average behavior of only those grains EMDs of which lie along the field direction. Therefore, the data from bulk pellets obtained in such small fields exhibit similar behavior as the \emph{single} grain measured using micro-Hall magnetometry. For larger external field, the signal due to all the randomly oriented grains with both easy and hard magnetization directions along the external field leads to a complicated average magnetic signal where the contribution of the AFM surface layer is suppressed. Additionally, an AFM phase is known to exhibit non-saturating behavior in $M$\,-\,$H$ measurements~\cite{BajpaiPRB2007, BentizPRB2011}. In the present case, the influence of the AFM surface layer is also reflected in the field sweep measurements where the magnetic signal increases linearly up to the measured field of 1\,T. The AFM surface layer around the FM micro-crystal may pin the surface FM spins, resulting in a non saturating slope in the FM saturation regime. However, we note here that we do not observe any systematic exchange bias effect which may be very small as a result of a much larger underlying FM CrO$_2$ layer and comparatively similar anisotropy energies. The values for $K_1$\,=\, 1.3\,-\,6\,$\times\,10^5$ erg/cm$^3$ and 2\,$\times\,10^5$\,erg/cm$^3$, respectively, are reported for CrO$_2$ and Cr$_2$O$_3$ in literature~\cite{DasAPL2010, DudkoPSS1971}.

Next, we discuss the analysis of stochastic Barkhausen jumps (see inset of Fig. \ref{Fig_Hyst_T} and Ref.~\cite{DasAPL2010}), due to the motion of a single DW, observed in the data obtained by sweeping the field at a slow rate of 1\,mT/min. A statistical analysis of these jumps in magnetic signal, particularly due to the motion of a \emph{single} DW, gives a reasonable picture of the distribution of pinning centers which characterize the potential landscape seen by the DW. In this respect, an analysis of the distribution of the jumps and their  temperature dependence is very useful to understand the nature and behavior of the pinning potentials.

In Fig.\,\ref{Fig_BH-jumps1}, we plot the sum of the amplitudes of all the individual Barkhausen jumps, observed at both grain ends, at different temperatures. Note that we measured the magnetic signal simultaneously at both ends and observed correlated Barkhausen jumps due to the motion of a single DW~\cite{DasAPL2010}. %We note here that the large non stochastic jump at $H^*$\,$\sim$\,20\,mT, which is not a Barkhausen jump and is not part of this analysis.
From the main panel of the figure, it is evident that the distance covered (i.e., the sum of the amplitudes of the individual jumps) by the DW by means of Barkhausen jumps decreases with increasing temperature following a pattern very similar to that observed for the total magnetic signal (cf. Fig.~\ref{Fig_Satfield_T}). %The large contribution of the spins of the AFM Cr$_2$O$_3$ layer at low temperatures (see Fig.\ref{Fig_Satfield_T}) may influence the effective external field acting on the DW.
Strikingly, the similarity in the temperature-dependent behavior is also evident in the \emph{total number of jumps} (which is ideally equal to the total number of pinning centers) varying as a function of temperature (see inset of Fig.~\ref{Fig_BH-jumps1}). These observations indicate that the AFM spins at the interface may play an important role in defining the potential landscape seen by the DW. Thus, this analysis is consistent with our observation that the magnetic behavior of CrO$_2$ is significantly influenced by the AFM surface layer.
%Although we see that the no. of possibly do not see the instrinsic pinning centers of the CrO$_2$ grain, it is clear that the magnitude of pinning potential is

In conclusion, we demonstrate here that the micro-Hall magnetometry can be used as a powerful method to extract information from measurements of a single magnetic entity which may otherwise be not evident from bulk measurements. From measurements of a \emph{single} micro-crystal of CrO$_2$ with a 2\,-\,5\,nm thin AFM surface layer of Cr$_2$O$_3$, we observe a robust signature of an unusual increase of the magnetic signal below $T\,\sim$\,60\,K\,$\ll\,T_C$. For polycrystalline bulk samples measured using SQUID magnetometry, the effect is observed only for measurements which are carried out in very small fields below $\sim$\,5\,mT. Comparison of the micro-Hall magnetometry data obtained from individual micro-crystals of CrO$_2$ with that of pure Cr$_2$O$_3$ indicates that the apparent increase in $M$ at low temperatures is influenced by the Cr$_2$O$_3$ surface layer. This underlines the important role played by the thin AFM layer on FM CrO$_2$ and should be taken in to account when analyzing the magneto-transport behavior of highly spin polarized CrO$_2$.

%\href{mailto:das@physik.uni-frankfurt.de}
\end{document}